\documentclass{article}
\usepackage{spconf}

\usepackage{cite}

  \usepackage[pdftex]{graphicx}

\usepackage{amsmath}
\usepackage{algorithm}
\usepackage{algorithmic}

  \usepackage[caption=false,font=footnotesize]{subfig}
\usepackage{url}

\usepackage{amsthm}

\newtheoremstyle{mydefinition}
{}
{}
{}
{0pt}
{\bfseries}
{.}
{ }
{\thmname{#1}\thmnumber{ #2}: \thmnote{#3}}

\theoremstyle{mydefinition}

\newtheoremstyle{myremark}
{}
{}
{}
{0pt}
{\bfseries}
{.}
{ }
{\thmname{#1}\thmnumber{ #2}: \thmnote{#3}}
\theoremstyle{myremark}

\newtheoremstyle{remarkshort}
{}
{}
{}
{0pt}
{\bfseries}
{.}
{ }
{\thmname{#1}\thmnumber{ #2}}

\theoremstyle{remarkshort}

\theoremstyle{remarkshort}

\newcommand{\comment}[1]{{}}

\usepackage{amssymb}
\usepackage{xspace}
\usepackage{bm}
\usepackage{bbm}

\let\originalleft\left
\let\originalright\right
\renewcommand{\left}{\mathopen{}\mathclose\bgroup\originalleft}
\renewcommand{\right}{\aftergroup\egroup\originalright}

\newcommand{\set}[1]{\ensuremath{\mathcal{#1}}\xspace} 
\newcommand{\mat}[1]{\ensuremath{\mathbf{#1}}\xspace} 
\renewcommand{\vec}[1]{\ensuremath{\mathbf{#1}}\xspace} 

\newcommand{\parens}[1]{{\left(#1\right)}\xspace}
\newcommand{\brackets}[1]{{\left[#1\right]}\xspace}
\newcommand{\braces}[1]{{\left\{#1\right\}}\xspace}

\newcommand{\doublebars}[1]{{\left\Vert#1\right\Vert}\xspace}

\newcommand{\complex}{\ensuremath{\mathbb{C}}\xspace}

\newcommand{\setcomplex}{\ensuremath{\complex}}

\newcommand{\setvector}[2]{\ensuremath{#1^{#2 \times 1}}\xspace}

\newcommand{\setvectorcomplex}[1]{\setvector{\setcomplex}{#1}}

\newcommand{\setmatrix}[3]{\ensuremath{#1^{#2 \times #3}}\xspace}

\newcommand{\setmatrixcomplex}[2]{\setmatrix{\setcomplex}{#1}{#2}}

\newcommand{\ctrans}{\ensuremath{^{{*}}}\xspace}

\newcommand{\logtwo}[1]{\ensuremath{\mathrm{log}_{2}\parens{#1}}}
\newcommand{\logten}[1]{\ensuremath{\mathrm{log}_{10}\parens{#1}}}

\newcommand{\pnorm}[2]{\ensuremath{\doublebars{#2}_{#1}}\xspace}

\newcommand{\normtwo}[1]{\pnorm{2}{#1}}

\newcommand{\distgauss}[2]{\ensuremath{\mathcal{N}\parens{#1,#2}}\xspace} 

\newcommand{\st}{\ensuremath{\mathrm{s.t.~}}\xspace}
\newcommand{\opt}{\ensuremath{^{\star}}\xspace}

\newcommand{\todB}[1]{\ensuremath{\brackets{#1}_{\mathrm{dB}}}}

\newcommand{\powernoise}{\ensuremath{P_{\mathrm{n}}}\xspace}

\newcommand{\powertx}{\ensuremath{P_{\mathrm{tx}}}\xspace}

\newcommand{\powertxue}{\ensuremath{\powertx^{\mathrm{UE}}}\xspace}
\newcommand{\powertxbs}{\ensuremath{\powertx^{\mathrm{BS}}}\xspace}

\newcommand{\snr}{\ensuremath{\mathsf{SNR}}\xspace}

\newcommand{\sinr}{\ensuremath{\mathsf{SINR}}\xspace}

\newcommand{\inr}{\ensuremath{\mathsf{INR}}\xspace}

\newcommand{\Nt}{\ensuremath{N_\mathrm{t}}\xspace} 
\newcommand{\Nr}{\ensuremath{N_\mathrm{r}}\xspace} 

\newcommand{\labeltx}{\mathrm{tx}}
\newcommand{\labelrx}{\mathrm{rx}}

\newcommand{\snrtx}{\ensuremath{\snr_{\labeltx}}\xspace}
\newcommand{\snrrx}{\ensuremath{\snr_{\labelrx}}\xspace}

\newcommand{\sinrtx}{\ensuremath{\sinr_{\labeltx}}\xspace}
\newcommand{\sinrrx}{\ensuremath{\sinr_{\labelrx}}\xspace}

\newcommand{\inrtx}{\ensuremath{\inr_{\labeltx}}\xspace}
\newcommand{\inrrx}{\ensuremath{\inr_{\labelrx}}\xspace}

\newcommand{\vhtx}{\vh_{\labeltx}\xspace}
\newcommand{\vhrx}{\vh_{\labelrx}\xspace}

\newcommand{\htx}{h_{\labeltx}\xspace}
\newcommand{\hrx}{h_{\labelrx}\xspace}

\newcommand{\Rtx}{R_{\labeltx}\xspace}
\newcommand{\Rrx}{R_{\labelrx}\xspace}

\def\vf{{\vec{f}}}

\def\vh{{\vec{h}}}

\def\vw{{\vec{w}}}

\def\vy{{\vec{y}}}

\def\mH{{\mat{H}}}

\def\sG{{\set{G}}}

\usepackage{xspace}
\usepackage[acronym,nogroupskip,nonumberlist,nopostdot]{glossaries}
\makeglossaries
\usepackage{relsize}

\usepackage{xcolor}

\newacronym{snr}{SNR}{signal-to-noise ratio}
\newacronym{sinr}{SINR}{signal-to-interference-plus-noise ratio}
\newacronym{inr}{INR}{interference-to-noise ratio}
\newacronym{sir}{SIR}{signal-to-interference ratio}
\newacronym{sqr}{SQR}{signal-to-quantization-noise ratio}
\newacronym{sqnr}{SQNR}{signal-to-quantization-plus-noise ratio}
\newacronym{ian}{IAN}{interference as noise}
\newacronym{ber}{BER}{bit error rate}
\newacronym{pn}{PN}{pseudorandom noise}
\newacronym{bfsk}{BFSK}{binary frequency shift keying}
\newacronym{fh}{FH}{frequency-hopped}
\newacronym{fh-bfsk}{FH-BFSK}{frequency-hopped binary frequency shift keying}
\newacronym{crc}{CRC}{cyclic redundancy check}
\newacronym{isi}{ISI}{intersymbol interference}
\newacronym{dsss}{DSSS}{direct-sequence spread spectrum}
\newacronym{ofdm}{OFDM}{orthogonal frequency-division multiplexing}
\newacronym{ofdma}{OFDMA}{orthogonal frequency-division multiple access}
\newacronym{sdr}{SDR}{software-defined radio}
\newacronym{tx}{TX}{transmitter}
\newacronym{rx}{RX}{receiver}
\newacronym{fdd}{FDD}{frequency-division duplexing}
\newacronym{tdd}{TDD}{time-division duplexing}
\newacronym{fdma}{FDMA}{frequency-division multiple access}
\newacronym{tdma}{TDMA}{time-division multiple access}
\newacronym{sdma}{SDMA}{space-division multiple access}
\newacronym[plural=MPCs]{mpc}{MPC}{multipath component}
\newacronym{mui}{MUI}{multi-user interference}
\newacronym{lsb}{LSB}{least significant bit}
\newacronym{jcas}{JCAS}{joint communication and sensing}

\newacronym{qam}{QAM}{quadrature amplitude modulation}
\newacronym{mqam}{MQAM}{M-ary quadrature amplitude modulation}

\newacronym{ls}{LS}{least-squares}
\newacronym{lms}{LMS}{least mean squares}
\newacronym{rls}{RLS}{recursive least-squares}
\newacronym{rzf}{RZF}{regularized zero-forcing}
\newacronym{mmse}{MMSE}{minimum mean square error}
\newacronym{lmmse}{LMMSE}{linear minimum mean square error}
\newacronym{mse}{MSE}{mean square error}
\newacronym{fft}{FFT}{fast Fourier transform}
\newacronym{dft}{DFT}{discrete Fourier transform}
\newacronym{dtft}{DTFT}{discrete-time Fourier transform}
\newacronym{ctft}{CTFT}{continuous-time Fourier transform}
\newacronym{ml}{ML}{machine learning}
\newacronym[plural=NNs]{nn}{NN}{neural network}
\newacronym[plural=RNNs]{rnn}{RNN}{recurrent neural network}
\newacronym[plural=ADCs]{adc}{ADC}{analog-to-digital converter}
\newacronym[plural=DACs]{dac}{DAC}{digital-to-analog converter}
\newacronym[plural=FPGAs]{fpga}{FPGA}{field-programmable gate array}
\newacronym{evm}{EVM}{error vector magnitude}
\newacronym{enob}{ENOB}{effective number of bits}
\newacronym{zf}{ZF}{zero-forcing}
\newacronym{rv}{r.v.}{random variable}
\newacronym{omp}{OMP}{orthogonal matching pursuit}
\newacronym{svd}{SVD}{singular value decomposition}

\newacronym{sdp}{SDP}{semidefinite programming}
\newacronym{psd}{PSD}{positive semidefinite}
\newacronym{nsd}{NSD}{negative semidefinite}

\newacronym{ks}{K-S}{Kolmogorov-Smirnov}

\newacronym{mad}{MAD}{median absolute deviation around the median}

\newacronym{agc}{AGC}{automatic gain control}
\newacronym{rf}{RF}{radio frequency}
\newacronym{if}{IF}{intermediate frequency}
\newacronym{los}{LOS}{line-of-sight}
\newacronym{nlos}{NLOS}{non-line-of-sight}
\newacronym{ple}{PLE}{path loss exponent}
\newacronym[plural=dB,firstplural=decibels (dB)]{db}{dB}{decibel}
\newacronym[plural=dBm,firstplural=decibel milliwatts (dBm)]{dbm}{dBm}{decibel milliwatts}
\newacronym{pa}{PA}{power amplifier}
\newacronym{lna}{LNA}{low noise amplifier}
\newacronym{vga}{VGA}{variable gain amplifier}
\newacronym{cw}{CW}{continuous wave}
\newacronym{papr}{PAPR}{peak-to-average power ratio}
\newacronym{usrp}{USRP}{Universal Software Radio Peripheral}
\newacronym{irr}{IRR}{image rejection ratio}
\newacronym{lo}{LO}{local oscillator}
\newacronym{vm}{VM}{vector modulator}
\newacronym{mmwave}{mmWave}{millimeter wave}
\newacronym{eirp}{EIRP}{effective isotropic radiated power}
\newacronym{rsrp}{RSRP}{reference signal received power}

\newacronym{csma}{CSMA}{carrier-sense multiple access}
\newacronym{csmaca}{CSMA/CA}{carrier-sense multiple access with collision avoidance}
\newacronym{csmacd}{CSMA/CD}{carrier-sense multiple access with collision detection}
\newacronym{mac}{MAC}{medium access control}
\newacronym{phy}{PHY}{physical layer}
\newacronym{4g}{4G}{fourth generation}
\newacronym{lte}{LTE}{Long-Term Evolution}
\newacronym{4glte}{4G LTE}{\gls{4g} \gls{lte}}
\newacronym{5g}{5G}{fifth generation}
\newacronym{nr}{NR}{New Radio}
\newacronym{5gnr}{5G NR}{5G New Radio}
\newacronym{ieee}{IEEE}{Institute of Electrical and Electronics Engineers}
\newacronym{wifi}{Wi-Fi}{IEEE 802.11}
\newacronym{lan}{LAN}{local area network}
\newacronym{wlan}{WLAN}{wireless local area network}
\newacronym[plural=BSs]{bs}{BS}{base station}
\newacronym[plural=SBSs]{sbs}{SBS}{small-cell base station}
\newacronym[plural=FD-SBSs]{fdsbs}{FD-SBS}{\gls{fd}-enabled \gls{sbs}}
\newacronym[plural=MBSs]{mbs}{MBS}{macrocell base station}
\newacronym[plural=UEs]{ue}{UE}{user equipment}
\newacronym{ul}{UL}{uplink}
\newacronym{dl}{DL}{downlink}
\newacronym{qos}{QoS}{Quality of Service}
\newacronym{fcc}{FCC}{Federal Communications Commission}
\newacronym{iab}{IAB}{integrated access and backhaul}
\newacronym{fab}{FAB}{fixed access and backhaul}
\newacronym{hetnet}{HetNet}{heterogeneous network}

\newacronym{siso}{SISO}{single-input single-output}
\newacronym{mimo}{MIMO}{multiple input multiple output}
\newacronym{sumimo}{SU-MIMO}{single-user \gls{mimo}}
\newacronym{mumimo}{MU-MIMO}{multi-user \gls{mimo}}
\newacronym{bf}{BF}{beamforming}
\newacronym{ca}{CA}{constant amplitude}
\newacronym{ula}{ULA}{uniform linear array}
\newacronym{upa}{UPA}{uniform planar array}
\newacronym[\glslongpluralkey={angles of arrival}]{aoa}{AoA}{angle of arrival}
\newacronym[\glslongpluralkey={angles of departure}]{aod}{AoD}{angle of departure}
\newacronym{dof}{DoF}{degrees of freedom}
\newacronym{csi}{CSI}{channel state information}
\newacronym{csit}{CSIT}{\gls{csi} at the transmitter}
\newacronym{csir}{CSIR}{\gls{csi} at the receiver}
\newacronym{cs}{CS}{compressed sensing}

\newacronym{fd}{FD}{in-band full-duplex}
\newacronym{hd}{HD}{half-duplex}
\newacronym{si}{SI}{self-interference}
\newacronym{sic}{SIC}{self-interference cancellation}
\newacronym{soi}{SoI}{signal of interest}
\newacronym{asic}{A-SIC}{analog \acrlong{sic}}
\newacronym{dsic}{D-SIC}{digital \gls{sic}}
\newacronym{star}{STAR}{simultaneous transmit and receive}
\newacronym{warp}{WARP}{Wireless Open-Access Research Platform}
\newacronym{bfc}{BFC}{beamforming cancellation}
\newacronym{ipi}{IPI}{inter-panel-interference}
\newacronym{ipic}{IPIC}{inter-panel-interference cancellation}

\newacronym{qcqp}{QCQP}{quadratically-constrained quadratic programming}
\newacronym{pdf}{PDF}{probability density function}
\newacronym{cdf}{CDF}{cumulative density function}
\newacronym{iid}{i.i.d.}{independently and identically distributed}

\newacronym{elf}{ELF}{extremely low frequency}
\newacronym{slf}{SLF}{super low frequency}
\newacronym{ulf}{ULF}{ultra low frequency}
\newacronym{vlf}{VLF}{very low frequency}
\newacronym{lf}{LF}{low frequency}
\newacronym{mf}{MF}{medium frequency}
\newacronym{hf}{HF}{high frequency}
\newacronym{vhf}{VHF}{very high frequency}
\newacronym{uhf}{UHF}{ultra high frequency}
\newacronym{shf}{SHF}{super high frequency}
\newacronym{ehf}{EHF}{extremely high frequency}
\newacronym{thf}{THF}{tremendously high frequency}

\newacronym{wncg}{WNCG}{Wireless Networking and Communications Group}
\newacronym{linc}{LINC}{Laboratory of Informatics, Networks, and Communications}
\newacronym{ut}{UT Austin}{The University of Texas at Austin}
\newacronym{uiuc}{UIUC}{University of Illinois at Urbana-Champaign}
\newacronym{usc}{USC}{University of Southern California}
\newacronym{mit}{MIT}{Massachusetts Institute of Technology}
\newacronym{berkeley}{UC Berkeley}{University of California, Berkeley}
\newacronym{osu}{OSU}{Ohio State University}

\newcommand{\mmwave}{\acrshort{mmwave}\xspace}
\newcommand{\mimo}{\gls{mimo}\xspace}

\newcommand{\bs}{\acrlong{bs}\xspace}
\newcommand{\bss}{\acrlongpl{bs}\xspace}
\newcommand{\gsnr}{\gls{snr}\xspace}
\newcommand{\ginr}{\gls{inr}\xspace}
\newcommand{\gsinr}{\gls{sinr}\xspace}

\newcommand{\figref}[1]{Fig.~\ref{#1}}

\usepackage{mathtools}

\usepackage{enumitem}

\definecolor{uclablue}{RGB}{39,116,174}
\definecolor{uclabluedarkest}{RGB}{0,59,92}
\definecolor{uclabluedarker}{RGB}{0,85,135}
\definecolor{uclabluelighter}{RGB}{139,184,232}
\definecolor{uclabluelightest}{RGB}{218,235,254}

\definecolor{uclagold}{RGB}{255,209,0}
\definecolor{uclagolddarker}{RGB}{255,199,44}
\definecolor{uclagolddarkest}{RGB}{255,184,28}

\definecolor{uclaviolet}{RGB}{130,55,165}
\definecolor{uclamagenta}{RGB}{255,0,165}

\usepackage{pgfplots}
\pgfplotsset{compat=newest}
\usetikzlibrary{plotmarks}
\usetikzlibrary{arrows.meta}
\usepgfplotslibrary{patchplots}
\usepackage{grffile}

\pgfplotsset{plot coordinates/math parser=false}

\begin{document}

%
\title{Active Beam Learning for Full-Duplex Wireless Systems}

%
%
%

\name{Jeong Min Kong and Ian P.~Roberts}
\address{%
Wireless Lab, Department of Electrical and Computer Engineering, UCLA\\
\{jeongminkong,ianroberts\}@ucla.edu%
}

\maketitle

\begin{abstract}
In this paper, we present a novel active beam learning method for in-band full-duplex wireless systems, that aims to design transmit and receive beams which suppress self-interference and maximize the sum spectral efficiency.
Rather than rely on explicit estimation of the downlink, uplink, and/or self-interference channels like in most existing work, our method instead actively probes all three channels through measurements of SNR and INR over a fixed number of time slots.
Then, once this probing concludes, all collected probing measurements are used to design transmit and receive beams which serve downlink and uplink in a full-duplex fashion.
We realize this active beam learning scheme through a network of LSTMs and DNNs, which learns to design each probing beam pair and subsequently extract and record valuable information from each probing measurement such that near-optimal serving beams can be designed following the probing stage. 
Simulation indicates that our method reliably suppresses self-interference while delivering near-maximal SNR on the downlink and uplink with merely 3--10 probing time slots, while exhibiting robustness to measurement noise and the structure of the self-interference channel.
\end{abstract}

\glsresetall

\section{Introduction} \label{sec:introduction}
Upgrading \bss with the ability to transmit downlink and receive uplink at the same time and over the same bandwidth---i.e., in-band full-duplex capability---can yield higher data rates, lower latency, broader coverage, and enhanced sensing capabilities, paving the way toward 6G networks \cite{smida_fd_6g_jsac_2023}.
The key to realizing full-duplex 6G networks lies in effectively mitigating the \textit{self-interference} inflicted by a given \bs's transmitter onto its own receiver, which would otherwise prohibitively degrade the uplink signal quality.
Analog and digital self-interference cancellation schemes have proved capable of realizing full-duplex in conventional low-frequency \bss \cite{singh_2020_acm,smida_fd_phy_2024}, but these schemes are less suitable in current 5G and emerging 6G radios, as they scale unfavorably to many antennas, wide bandwidths, and high carrier frequencies \cite{smida_fd_6g_jsac_2023,roberts_wcm}.
Motivated by this, recent works \cite{roberts_lonestar,lopez_analog_2022,satyanarayana_hybrid_2019,roberts_steer,roberts_realworld} have harnessed beamforming to cancel the self-interference coupled between the transmit and receive antenna arrays of a massive \mimo or millimeter wave (mmWave) wireless system.
These schemes have proven capable of reducing self-interference to below the noise floor and are thus a promising route to unlocking full-duplex wireless systems but exhibit noteworthy practical shortcomings.

Most notably, almost all existing full-duplex beam designs rely on explicit estimation of the users' downlink and uplink channels as well as the \mimo self-interference channel.
In massive MIMO and \mmwave systems, the number of antennas can be on the order of dozens or even hundreds, and estimating these high-dimensional channels would thus be resource-expensive and impractical.
This is evident even in today's half-duplex massive \mimo and \mmwave systems, which circumvent explicit estimation of downlink and uplink channels via codebook-based beam alignment procedures \cite{ethan_beam}.
To our knowledge, the full-duplex beam designs in \cite{roberts_steer,roberts_realworld} are the only ones which do not rely on explicit channel estimation but rather power measurements across the downlink, uplink, and/or self-interference channels.
While experimentally validated, these methods in \cite{roberts_steer,roberts_realworld} require around 50--500 measurements per downlink-uplink user pair, which may make them unsuitable in 6G networks, given their resource constraints.

In this paper, we propose a novel active beam learning method to design near-optimal transmit and receive beams of a full-duplex \bs, without explicit channel estimation and with only a few measurements.  
Our approach involves the \bs first jointly probing the downlink, uplink, and self-interference channels over a \textit{fixed} number of time slots, from which it extracts implicit information about all three channels to design the final serving beams. 
Structured as a sequential decision-making process, the \bs \textit{actively} designs the probing beams in each time slot based on its prior probing measurements, allowing it to tailor each probing sequence to the particular user pair and self-interference channel realization.
Notably, this method bypasses any explicit channel estimation and instead relies solely on power measurements to design effective probing and serving beams.
Simulation results demonstrate that with only 3--10 probing time slots, our scheme can deliver near-maximal spectral efficiency on the downlink and uplink.

\section{System Model} \label{sec:system-model}

\begin{figure}[!t]
    \centering
    \includegraphics[width=\linewidth,height=0.22\textheight,keepaspectratio]{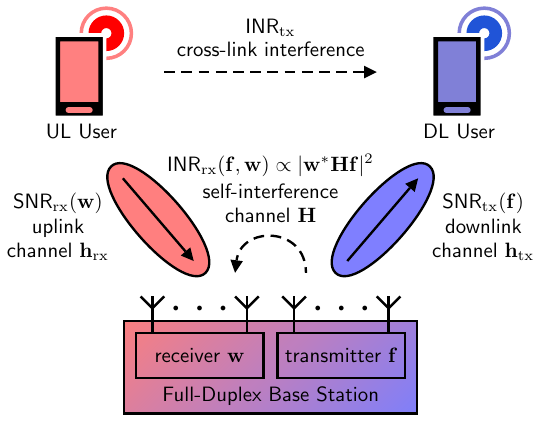}
    \caption{A full-duplex base station transmits downlink to one user while receiving uplink from another user at the same time and same frequency.}
    \label{fig:system}
\end{figure}

In this work, we consider a base station serving a single-antenna downlink user and a single-antenna uplink user simultaneously over the same frequency, as depicted in \figref{fig:system}.
The base station utilizes two separate antenna arrays, one for transmission and the other for reception; the transmit array has $\Nt$ antennas and analog beamforming weights $\vf \in \setvectorcomplex{\Nt}$, and the receive array has $\Nr$ antennas and analog beamforming weights $\vw \in \setvectorcomplex{\Nr}$. 
The \glspl{snr} of the downlink and uplink are
\begin{align}
    \snrtx(\vf) &= \frac{\powertxbs \cdot |\vhtx\ctrans\vf|^2}{\powernoise^{\mathrm{UE}}} \\
    \snrrx(\vw) &= \frac{\powertxue \cdot |\vw\ctrans\vhrx|^2}{\powernoise^{\mathrm{BS}}},
\end{align}
where $\powertxbs$ is the transmit power of the base station, $\powernoise^{\mathrm{UE}}$ is the noise power at the downlink user, $\vec{\htx} \in \setmatrixcomplex{\Nt}{1}$ is the downlink channel, $\powertxue$ is the transmit power of the uplink user, $\powernoise^{\mathrm{BS}}$ is the noise power at the base station, and $\vec{\hrx} \in \setmatrixcomplex{\Nr}{1}$ is the uplink channel.

Since the uplink and downlink are operating at the same frequency band, the base station's transmit array inflicts so-called self-interference upon its own receive array across the \mimo channel $\mH \in \setmatrixcomplex{\Nr}{\Nt}$.
While still the topic of active research, a plausible model for the self-interference channel $\mH$, backed by measurements \cite{roberts_realworld}, is a Rician fading model
\begin{align}\label{eq:si}
    \mH = \sqrt{\frac{\kappa}{\kappa + 1}} \, \bar\mH + \sqrt{\frac{1}{\kappa + 1}} \, \tilde\mH.
\end{align}
Here, $\kappa$ is the Rician factor, $\bar\mH$ is the static part of the self-interference channel caused by near-field coupling between the antenna arrays, and $\tilde\mH$ is a time-varying component that stems from unpredictable environmental factors such as reflections.
The strength of self-interference coupled by a given transmit beam $\vf$ and receive beam $\vw$ can be captured by the \gls{inr} given by
\begin{align}
    \inrrx(\vf,\vw) = \frac{\powertxbs \cdot |\vw\ctrans \mat{H} \vf|^2}{\powernoise^{\mathrm{BS}}}.
\end{align}
Note that there is also interference induced by the uplink user on the downlink user, as shown in \figref{fig:system}; the INR of this cross-link interference will be denoted as $\mathsf{INR}_{\mathrm{tx}}$. 
While self-interference depends on the transmit and receive beams at the base station, the cross-link interference depends only on the users and the channel between them.

Downlink and uplink \glspl{sinr}, which account for cross-link interference and self-interference, respectively, can be expressed as
\begin{align}
    \sinrtx(\vf) &= \frac{\snrtx(\vf)}{1+\inrtx} \\
    \sinrrx(\vf,\vw) &= \frac{\snrrx(\vw)}{1+\inrrx(\vf,\vw)}.
\end{align}
The achievable downlink, uplink, and sum spectral efficiencies, with transmit beam $\vf$ and receive beam $\vw$, are then
\begin{align}
\Rtx(\vf) &= \logtwo{1+\sinrtx(\vf)} \\
\Rrx(\vf,\vw) &= \logtwo{1+\sinrrx(\vf,\vw)} \\
R(\vf,\vw) &= \Rtx(\vf) + \Rrx(\vf,\vw).
\end{align}

\section{Active Beam Learning for Full-Duplex} \label{sec:contribution}

\begin{figure}
    \centering
    \includegraphics[width=\linewidth,height=\textheight,keepaspectratio]{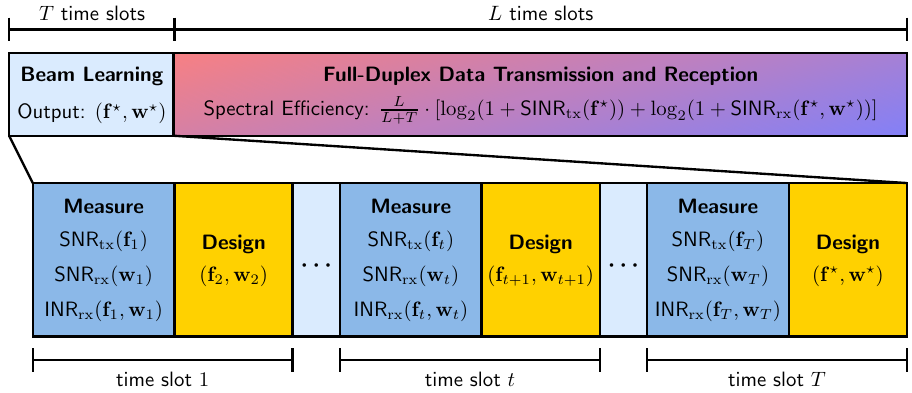}
    \caption{Our envisioned active beam learning solution.}
    \label{fig:outline}
\end{figure}

\begin{figure*}
    \centering
    \includegraphics[width=\linewidth,height=\textheight,keepaspectratio]{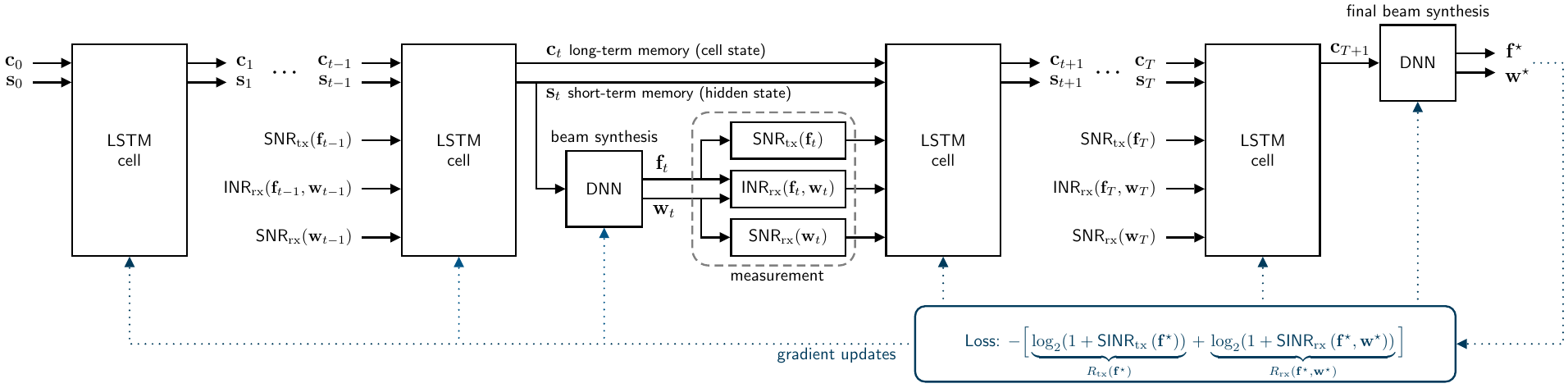}
    \caption{An LSTM-based implementation of our proposed active beam learning solution for full-duplex wireless systems.}
    \label{fig:LSTM}
\end{figure*}

In this section, we aim to design $\vf$ and $\vw$ in order to maximize the sum spectral efficiency $R(\vf,\vw)$, without assuming knowledge of the downlink, uplink, or self-interference channels \textit{a priori}.
As mentioned in the introduction, most prior approaches to this problem assumes these channels have been reliably estimated beforehand and thus ignores the resources consumed by such estimation.
In contrast, we consider a resource-constrained setting, wherein the base station is allotted a \textit{fixed} number of time slots to take measurements of downlink, uplink, and self-interference before designing the beams used for full-duplex data transmission and reception.
Constraining our design to a fixed overhead will thus make it more readily adopted in real-world wireless networks.

Our envisioned approach to this problem is shown in \figref{fig:outline}, where the base station is allotted $T$ time slots before designing the beams it uses to serve downlink and uplink for $L$ time slots.
In the first time slot, the proposed \textit{beam learning} process begins with the base station measuring $\snrtx(\vf_1)$, $\snrrx(\vw_1)$, and $\inrrx(\vf_1,\vw_1)$ using some initial \textit{probing} beam pair $(\vf_1,\vw_1)$, followed by designing the next probing beam pair $(\vf_2,\vw_2)$ based on these initial measurements. 
In the next time slot, the base station uses these new probing beams to measure the SNRs and the INR, and then similarly generates the next probing beams $(\vf_3,\vw_3)$ based on \textit{all} prior measurements. 
This process is repeated for all $T$ probing time slots, after which the base station uses information gathered from the entire probing stage to design the \emph{serving} transmit and receive beams $(\vf\opt,\vw\opt)$, which ideally maximizes the sum spectral efficiency. 
Designing the sequence of probing beams $\braces{(\vf_t,\vw_t)}_{t=1}^{T}$ in the 
aforementioned \textit{active} manner \cite{active_sensing_2021} will prove capable of designing near-optimal $(\vf\opt,\vw\opt)$, even when the probing overhead $T$ is relatively small.

\subsection{Problem Formulation}
Inspired by the work of \cite{active_sensing_2021}, our envisioned active beam learning problem, as just described and illustrated in \figref{fig:outline}, can be formulated as
\begin{subequations} \label{eq:problem}
\begin{align}
\max_{\substack{\{\mathcal{F}_t (\cdot)\}_{t=0}^{T-1},\\\mathcal{G} (\cdot)}} \ & 
\Rtx(\vf\opt) + \Rrx(\vf\opt,\vw\opt) \\
\st \ 
&(\vf_{t+1},\vw_{t+1}) = \mathcal{F}_t(\vf_{1:t},\vw_{1:t},\vy_{1:t}) \ \forall \ t \\ 
&(\vf\opt,\vw\opt) = \mathcal{G}\parens{\vf_{1:T},\vw_{1:T},\vy_{1:T}} \\
& \normtwo{\vf_t}^2 = \normtwo{\vf\opt}^2 = \Nt \ \forall \ t = 1, \dots, T \label{eq:power-const-f} \\
& \normtwo{\vw_t}^2 = \normtwo{\vw\opt}^2 = \Nr \ \forall \ t = 1, \dots, T, \label{eq:power-const-w}
\end{align}
\end{subequations}
where $\vy_t = \brackets{\snrtx(\vf_t), \snrrx(\vw_t), \inrrx(\vf_t,\vw_t)}$ are the measurements taken with probing beams $\vf_{t}$ and $\vw_{t}$ during time slot $t$. 
Our goal in this problem is to find the series of functions $\{\mathcal{F}_t (\cdot)\}_{t=0}^{T-1}$ and the function $\mathcal{G} (\cdot)$ that produces a final serving beam pair $(\vf\opt,\vw\opt)$ which maximizes the sum spectral efficiency. 
Here, $\mathcal{F}_t(\cdot)$ is a function that outputs the probing beams for time slot $t+1$ based on all measurements collected through time $t$, whereas $\mathcal{G}(\cdot)$ is a function that designs the final serving beams based on all $T$ measurements from the probing phase.
Note that the same functions $\{\mathcal{F}_t (\cdot)\}_{t=0}^{T-1}$ and $\sG(\cdot)$ are to be used across user pairs and self-interference channel realizations. 

By developing an \emph{active} probing strategy in this fashion, our aim is to collect a unique sequence of measurements tailored specifically to the user pair being served. 
This will allow the probing sequence to implicitly account for the structure of the self-interference channel in relation to those users' downlink and uplink channels.
In turn, each probing sequence will be more targeted to the particular user pair and self-interference channel realization (than a non-active approach)
and will thus give our active beam learning solution the potential to excel with only a few probing measurements.
It is also worth noting that our proposed design is based solely on noncoherent SNR and INR measurements, lending itself well to certain practical deployments.

\subsection{LSTM-Based Active Beam Learning}

To implement our active beam learning solution and (approximately) solve problem \eqref{eq:problem}, we adopt the long short-term memory (LSTM) architecture illustrated in \figref{fig:LSTM}, taking inspiration from \cite{active_sensing_2021}.
LSTMs are especially well-suited for our active beam learning task, as they can effectively learn to extract valuable information from each probing measurement and retain a record of this extracted information, allowing us to leverage all prior probing measurements when designing subsequent probing beams or the final serving beams.
In the remainder of this section, we will explain our LSTM-based approach and describe how the architecture shown in \figref{fig:LSTM} is trained to solve problem \eqref{eq:problem}.

As shown in \figref{fig:LSTM}, each LSTM cell indexed at time slot $t$ takes as input the following three sets of parameters: 
(i) the downlink SNR, uplink SNR, and INR measured with probing beams $\vf_{t}$ and $\vw_{t}$; 
(ii) the LSTM hidden state at time $t$, denoted by $\mathbf{s}_{t}$; 
and (iii) the LSTM cell state at time $t$, denoted by $\mathbf{c}_{t}$. 
The LSTM cell then outputs the updated hidden and cell states, denoted as $\mathbf{s}_{t+1}$ and $\mathbf{c}_{t+1}$, respectively, and feeds the new hidden state into a deep neural network (DNN) to synthesize probing beams $(\vf_{t+1},\vw_{t+1})$ for use in the next time slot, $t + 1$. 
As the hidden state acts as a memory that carries relevant information from both recent and past measurements, the described procedure synthesizes probing beams $(\vf_{t+1},\vw_{t+1})$ based on all prior measurements. 
The aforementioned process is repeated sequentially across $T-1$ time slots. 
Following the final measurements at time slot $T$, the last cell state $\mathbf{c}_{T+1}$, which functions as a long-term memory holding valuable information gathered throughout the entire probing phase, is input to another DNN to synthesize the serving beams $\vf\opt$ and $\vw\opt$. 
Assuming we have models or datasets of the downlink, uplink, and self-interference channels, the model parameters can be optimized to maximize the sum spectral efficiency through backpropagation.
Then, after training, the model can be deployed to actively generate probing beams for a given user pair and self-interference channel realization and, based on these probing measurements, can output a final beam pair $(\vf\opt,\vw\opt)$ for serving downlink and uplink in a full-duplex fashion, realizing the solution envisioned in \figref{fig:outline}.

\section{Evaluation} \label{sec:numerical-results}

To evaluate our method, we consider a $28$~GHz full-duplex base station equipped with two $8$-element half-wavelength linear arrays spaced apart horizontally $10 \lambda$, which it uses to serve single-antenna, line-of-sight users located uniformly between $-60^\circ$ and $60^\circ$ in azimuth. 
We model the static component of the self-interference channel $\bar\mH$ with the near-field spherical-wave model \cite{spherical_2005} and the time-varying component $\tilde\mH$ as a Rayleigh fading channel drawn independently from user pair to user pair. 
To account for measurement errors that inevitably arise in practice, we assume (during training and evaluation) that the probing measurements (in dB) are corrupted by zero-mean Gaussian noise with variance $\sigma^2$.\footnote{Specifically, we assume all \gsnr and \ginr measurements are of the form $\todB{\hat{\snr}_\labeltx(\vf_t)} = \todB{\snrtx(\vf_t)} + \distgauss{0}{\sigma^2}$, where $\distgauss{0}{\sigma^2}$ is Gaussian with mean zero and variance $\sigma^2$ and $\todB{x} = 10\,\logten{x}$.}
To satisfy the power constraints of \eqref{eq:power-const-f} and \eqref{eq:power-const-w}, we perform appropriate normalizations to $\vf_{t}$ and $\vw_{t}$ at each time slot. 
Additionally, we normalize $\vhtx$, $\vhrx$, and $\mH$ such that the maximum achievable downlink and uplink SNRs are $10$~dB and the maximum possible INR is $40$~dB. 
Lastly, for simplicity, we assume that there is no cross-link interference present between the downlink and uplink users. 

We use TensorFlow to implement our LSTM-based active beam learning solution. 
The dimensions of the LSTM hidden and cell states, $\mathbf{s}_{t}$ and $\mathbf{c}_{t}$, are set to $512$. The probing beam synthesis and final beam synthesis DNNs comprise of three hidden layers, each containing 1024 neurons with ReLU activation. 
Note that the output dimension of each of these DNNs is $2 \times (\Nt + \Nr)$ in order to account for both the real and imaginary components of the complex beamforming weights. 
During training, we utilize Adam optimizer with a learning rate of $10^{-4}$ and a mini-batch of size $128$. 
As described, we set the loss function to $-R(\vf\opt,\vw\opt)$ to optimize the model to maximize the sum spectral efficiency.

\begin{figure}[!t]
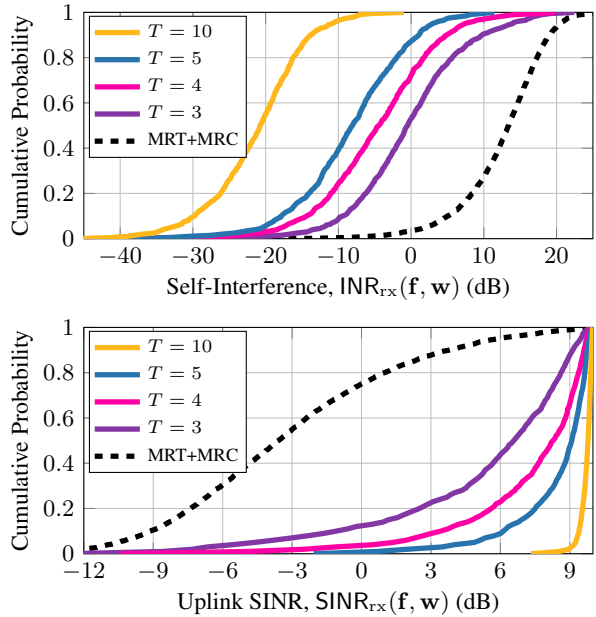
%
    \centering%
    \input{plots/CDF_INRRx.tex}%
    
    \input{plots/CDF_SINRRx}%
    
    \caption{Empirical CDFs of $\inrrx$ and $\sinrrx$ over $1000$ random user pairs, when $\sigma^2 = 0.4$ and $\kappa = 7$ dB.}%
    \label{fig:inr-sinr-rate-cdf}%
\end{figure}%

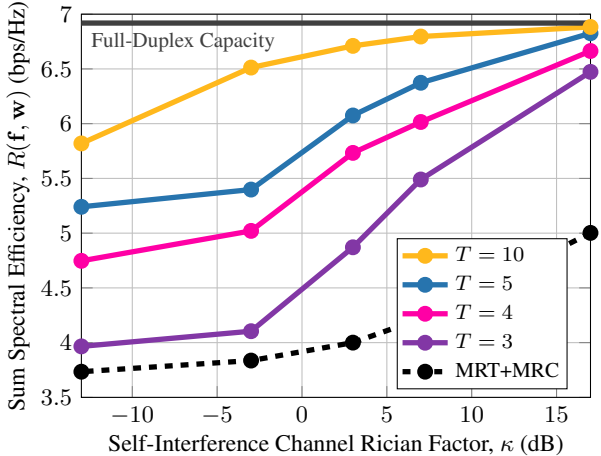
\begin{figure}[t]
    \centering%
%
%
\definecolor{mycolor1}{rgb}{1.00000,0.00000,1.00000}%
\definecolor{mycolor2}{rgb}{0.30100,0.74500,0.93300}%
\begin{tikzpicture}

\begin{axis}[%
width=\linewidth,
height=0.8\linewidth,
xmin=-13,
xmax=17,
xlabel style={font=\color{white!0!black}\small},
xlabel={Self-Interference Channel Rician Factor, $\kappa$ (dB)},
ymin=3.5,
ymax=7,
ylabel style={font=\color{white!0!black}\small},
ylabel={Sum Spectral Efficiency, $R(\vf,\vw)$ (bps/Hz)},
axis background/.style={fill=white},
xmajorgrids,
ymajorgrids,
xtick distance=5,
ytick distance=0.5,
ticklabel style={font=\color{white!0!black}\small},
xlabel shift = -3pt,
ylabel shift = -3pt,
legend style={at={(0.955,0.021)}, anchor=south east, legend cell align=left, align=left, draw=white!0!black, font=\footnotesize, inner sep=1pt, reverse legend},
]

\addplot [color=black, dashed,  line width=2.0pt, mark=*, mark options={solid, black}]
table[row sep=crcr]{%
    17	5.0024\\
    7	4.2188\\
    3	4.0001\\
    -3	3.8362\\
    -13	3.7344\\
};
\addlegendentry{MRT+MRC}

\addplot [color=uclaviolet, line width=2.0pt, mark=*, mark options={solid, uclaviolet}]
  table[row sep=crcr]{%
17	6.4731\\
7	5.49\\
3	4.8709\\
-3	4.1045\\
-13	3.9658\\
};
\addlegendentry{$T = 3$}

\addplot [color=uclamagenta, line width=2.0pt, mark=*, mark options={solid, uclamagenta}]
  table[row sep=crcr]{%
17	6.6637\\
7	6.0153\\
3	5.7331\\
-3	5.0207\\
-13	4.7471\\
};
\addlegendentry{$T = 4$}

\addplot [color=uclablue, line width=2.0pt, mark=*, mark options={solid, uclablue}]
  table[row sep=crcr]{%
17	6.8259\\
7	6.3735\\
3	6.0748\\
-3	5.3981\\
-13	5.2406\\
};
\addlegendentry{$T = 5$}

\addplot [color=uclagolddarkest, line width=2.0pt, mark=*, mark options={solid, uclagolddarkest}]
  table[row sep=crcr]{%
17	6.8834\\
7	6.7959\\
3	6.7105\\
-3	6.5124\\
-13	5.8194\\
};
\addlegendentry{$T = 10$}

\addplot [color=darkgray, line width=2.0pt, mark=p, mark options={solid, black}, forget plot]
table[row sep=crcr]{%
    17	6.91886\\
    7	6.91886\\
    3	6.91886\\
    -3	6.91886\\
    -13	6.91886\\
} node[below=-0.3mm,pos=0.8] {\footnotesize Full-Duplex Capacity};

\end{axis}
\end{tikzpicture}%
    \caption{Sum spectral efficiency as a function of $\kappa$ (averaged over 1000 random user pairs), when $\sigma^2 = 0.4$.}
    \label{fig:kappa_vs_rsum}
\end{figure}

We compare our approach against two baselines: (i) maximum ratio transmission (MRT) plus maximum ratio combining (MRC), which steers beams directly toward each user to maximize their SNRs, but ignores self-interference, and (ii) the full-duplex capacity, achieved by MRT+MRC in the absence of self-interference. 
\figref{fig:inr-sinr-rate-cdf} presents the empirical cumulative distribution functions (CDFs) of $\inrrx$ and $\sinrrx$ over $1000$ random user pairs, when the measurement noise variance is $\sigma^2 = 0.4$ and the self-interference channel Rician factor is $\kappa = 7$~dB. 
The first key observation to make is that self-interference is often 10--20~dB above the noise floor with MRT+MRC, making it unsuitable for full-duplex operation.
In contrast, active beam learning is able to design serving beams $(\vf\opt,\vw\opt)$ that are far more robust to self-interference with only 3--10 time slots for probing.
With $T=10$, remarkably all 1000 randomly generated user pairs enjoy self-interference below the noise floor.
This translates to higher uplink \glspl{sinr}, which closely hugs the upper bound of 10~dB as the allotted probing time is increased to $T=10$.
Although MRT+MRC maximizes \gsnr, its high self-interference leads to poor uplink \gsinr.

In \figref{fig:kappa_vs_rsum}, we further investigated the impact of the structure of the self-interference channel $\mH$ on performance. 
Fixing $\sigma^2=0.4$, we plot the sum spectral efficiency as a function of the self-interference channel Rician factor $\kappa$, averaged over $1000$ random user pairs. 
Recalling \eqref{eq:si}, increasing $\kappa$ results in a more deterministic, spherical-wave self-interference channel, while a lower value leads to a more random, Gaussian channel.
From the plot, it is evident that our method achieves higher sum spectral efficiency as $\kappa$ increases; we attribute this to two main reasons. 
First, from the dashed MRT+MRC curve, we can see that a spherical-wave channel is structurally more favorable on average than a Gaussian one, also observed in \cite{roberts_lonestar}.
Second, the performance of active beam learning increases with $\kappa$ due to the simple fact that the channel becomes more deterministic, making it easier to learn probing strategies and final synthesis of $(\vf\opt,\vw\opt)$. 
With that being said, even when $\kappa$ is low and $\mH$ is heavily Gaussian, active beam learning can still provide performance approaching the full-duplex capacity with sufficient probing time $T$.
Altogether, these numerical results suggest that active beam learning is capable of intelligently probing to strategically design serving beams $(\vf\opt,\vw\opt)$ which suppress self-interference without compromising downlink and uplink SNR.

\section{Conclusion and Future Work} \label{sec:conclusion}

This work introduced an active beam learning scheme that aims to maximize the sum spectral efficiency of a full-duplex system.
This is realized using a network of LSTMs and DNNs which learns to actively probe the downlink, uplink, and self-interference channels, rather than explicitly estimate them.  
We demonstrated that our approach is able to suppress self-interference to below noise and attain sum spectral efficiencies that approach the full-duplex capacity when the probing time is sufficiently large.
Our results also illustrate its robustness to measurement noise and to the structure of self-interference. 
Valuable future work may adapt our approach to scale more favorably to many users or to exploit temporal or spatial correlations in the system.

\bibliographystyle{bibtex/IEEEtran}
{\small \bibliography{bibtex/IEEEabrv,refs}}

\begin{thebibliography}{10}
\providecommand{\url}[1]{#1}
\csname url@samestyle\endcsname
\providecommand{\newblock}{\relax}
\providecommand{\bibinfo}[2]{#2}
\providecommand{\BIBentrySTDinterwordspacing}{\spaceskip=0pt\relax}
\providecommand{\BIBentryALTinterwordstretchfactor}{4}
\providecommand{\BIBentryALTinterwordspacing}{\spaceskip=\fontdimen2\font plus
\BIBentryALTinterwordstretchfactor\fontdimen3\font minus
  \fontdimen4\font\relax}
\providecommand{\BIBforeignlanguage}[2]{{%
\expandafter\ifx\csname l@#1\endcsname\relax
\typeout{** WARNING: IEEEtran.bst: No hyphenation pattern has been}%
\typeout{** loaded for the language `#1'. Using the pattern for}%
\typeout{** the default language instead.}%
\else
\language=\csname l@#1\endcsname
\fi
#2}}
\providecommand{\BIBdecl}{\relax}
\BIBdecl

\bibitem{smida_fd_6g_jsac_2023}
B.~Smida, A.~Sabharwal, G.~Fodor, G.~C. Alexandropoulos, H.~A. Suraweera, and
  C.-B. Chae, ``Full-duplex wireless for {6G}: Progress brings new
  opportunities and challenges,'' \emph{{IEEE} J. Sel. Areas Commun.}, vol.~41,
  no.~9, pp. 2729--2750, Sep. 2023.

\bibitem{singh_2020_acm}
V.~Singh, S.~Mondal, A.~Gadre, M.~Srivastava, J.~Paramesh, and S.~Kumar,
  ``Millimeter-wave full duplex radios,'' in \emph{Proc. {ACM} MobiCom}, Apr.
  2020.

\bibitem{smida_fd_phy_2024}
B.~Smida, R.~Wichman, K.~E. Kolodziej, H.~A. Suraweera, T.~Riihonen, and
  A.~Sabharwal, ``In-band full-duplex: The physical layer,'' \emph{Proc.
  {IEEE}}, vol. 112, no.~5, pp. 433--462, May 2024.

\bibitem{roberts_wcm}
I.~P. Roberts, J.~G. Andrews, H.~B. Jain, and S.~Vishwanath, ``Millimeter-wave
  full duplex radios: New challenges and techniques,'' \emph{{IEEE} Wireless
  Commun.}, vol.~28, no.~1, pp. 36--43, Feb. 2021.

\bibitem{roberts_lonestar}
I.~P. Roberts, S.~Vishwanath, and J.~G. Andrews, ``\textsc{LoneSTAR}: Analog
  beamforming codebooks for full-duplex millimeter wave systems,'' \emph{{IEEE}
  Trans. Wireless Commun.}, vol.~22, no.~9, pp. 5754--5769, Sep. 2023.

\bibitem{lopez_analog_2022}
R.~L\'opez-Valcarce and M.~Mart\'inez-Cotelo, ``Full-duplex {mmWave} {MIMO}
  with finite-resolution phase shifters,'' \emph{{IEEE} Trans. Wireless
  Commun.}, vol.~21, no.~11, pp. 8979--8994, Nov. 2022.

\bibitem{satyanarayana_hybrid_2019}
K.~Satyanarayana, M.~El-Hajjar, P.~Kuo, A.~Mourad, and L.~Hanzo, ``Hybrid
  beamforming design for full-duplex millimeter wave communication,''
  \emph{{IEEE} Trans. Veh. Technol.}, pp. 1394--1404, Feb. 2019.

\bibitem{roberts_steer}
I.~P. {Roberts}, A.~Chopra, T.~Novlan, S.~Vishwanath, and J.~G. {Andrews},
  ``\textsc{Steer}: Beam selection for full-duplex millimeter wave
  communication systems,'' \emph{{IEEE} Trans. Commun.}, vol.~70, no.~10, pp.
  6902--6917, Oct. 2022.

\bibitem{roberts_realworld}
I.~P. Roberts, Y.~Zhang, T.~Osman, and A.~Alkhateeb, ``Real-world evaluation of
  full-duplex millimeter wave communication systems,'' \emph{{IEEE} Trans.
  Wireless Commun.}, vol.~23, no.~9, pp. 10\,803--10\,819, Sep. 2024.

\bibitem{ethan_beam}
Y.~Heng \emph{et~al.}, ``Six key challenges for beam management in {5.5G} and
  {6G} systems,'' \emph{{IEEE} Commun. Mag.}, vol.~59, no.~7, pp. 74--79, Jul.
  2021.

\bibitem{active_sensing_2021}
F.~Sohrabi, T.~Jiang, W.~Cui, and W.~Yu, ``Active sensing for communications by
  learning,'' \emph{{IEEE} J. Sel. Areas Commun.}, vol.~40, no.~6, pp.
  1780--1794, Jun. 2022.

\bibitem{spherical_2005}
J.-S. Jiang and M.~A. Ingram, ``Spherical-wave model for short-range {MIMO},''
  \emph{{IEEE} Trans. Commun.}, vol.~53, no.~9, pp. 1534--1541, Sep. 2005.

\end{thebibliography}

\end{document}